# Studies of superconductivity in $U(Pt_{1-x}Pd_x)_3$ for $x < 0.006$

M. J. Graf [a], R. J. Keizer[b], A. de Visser[b], and J. J. M. Franse[b]

a - Department of Physics, Boston College, Chestnut Hill, MA 02167, USA

b - Van der Waals - Zeeman Inst., University of Amsterdam, 1018 XE Amsterdam, The Netherlands

We report measurements of the specific heat and resistivity ($T < 1$ K) for high quality polycrystals of $U(Pt_{1-x}Pd_x)_3$ with $x < 0.006$. The $T_c$-x phase diagram can be constructed, and superconductivity is destroyed for $x \approx 0.006$; this is approximately the same concentration above which the onset of large-moment antiferromagnetism is observed to occur. The splitting of the double superconducting transition increases smoothly with increasing Pd content, and is large enough that for Pd concentrations $0.004 < x < 0.006$ only the superconducting A-phase will be present.

Corresponding author:    Michael J. Graf
Department of Physics
Boston College
Chestnut Hill, MA 02167 USA
E-mail: GRAFM@BC.EDU
Fax: (+001) 617 552 8478

Pd-substitution is a powerful method to study unconventional superconductivity and magnetism in $UPt_3$. Studies of $U(Pt_{1-x}Pd_x)_3$ for $x \leq 0.002$ showed that Pd-substitution is unique in that it increases the splitting $\Delta T_c$ between the A and B superconducting phases by 40 mK for x=0.002 [1]. Addition of Pd also increases the zero-temperature moment associated with "small-moment" antiferromagnetism (SMAF) above the $x = 0$ value of 0.02 $\mu_B$, and the correlation between the increased moment and $\Delta T_c$ has been confirmed [2]. For $x > 0.01$, "large-moment" antiferromagnetism (LMAF) is observed[3] through μSR[4] and other conventional methods; in contrast, SMAF is convincingly observed only via neutron and magnetic x-ray scattering [5]. In this work we address two questions: does the splitting increase with continued increase of Pd above $x = 0.002$, and at what concentration does the superconducting $T_c$ approach 0 K?

Polycrystalline samples with $x = 0$, 0.0025, 0.003, 0.0035, 0.004, and 0.005 were studied. For x=0, the ratio of the RT resistivity to the extrapolated T=0 K normal-state resistivity $\rho_o$ is over 1000. $\rho_o$ varies linearly with x, showing that the series is uniform. In Fig. 1 we show the resistive $T_{cA}$ versus x (solid symbols); $T_{cA}$ approaches 0 K at $x = 0.006$. This is also the estimated concentration at which the LMAF state begins, based on neutron[3] and μSR[4] studies for $x \geq 0.005$. This strongly suggests that the LMAF and superconducting phases compete with one another, presumably because the superconducting pairing mechanism is based on AFM spin-fluctuations. We are investigating several samples with $0.006 < x < 0.010$ to study the phase diagram in more detail.

In Fig. 2 we show the specific heat results. The splitting clearly increases smoothly with Pd-concentration (see also Fig. 1, open symbols). $\Delta T_c$ is increased to 150 mK for x=0.003, compared to 55 mK for $x = 0$. Comparing the variation of $T_{cA}$ and $\Delta T_c$ with x, we estimate that for x > 0.004 there will be only one superconducting phase (A). We also see that for x ≥ 0.0025 $\Delta C/T(T_{cB}) < \Delta C/T(T_{cA})$, where $\Delta C/T(T_{ci})$ is the specific heat discontinuity at $T_{ci}$ measured relative to the normal state. This appears to violate the stability condition derived from a Ginzburg-Landau analysis near $T_c$ within the "E-rep" models of superconductivity [6]. However, $\Delta T_c$ is large, and there is a substantial temperature variation of C/T for $T_{cB} < T < T_{cA}$. Since the calculations are carried out only to fourth-order in the order parameter, they constrain C/T to be constant, and the comparison is invalid. However, correcting for this temperature dependence of C/T by using the $\Delta C/T(T_{cB})$ as measured from the A-phase value, we estimate that the ratio of the discontinuities is nearly constant and has a magnitude consistent with the weak-coupling E-rep models [6]. A more exact analysis of the data will require calculations which are carried out to higher order.

Our results demonstrate that the temperature-dependent properties of the A-phase can now be studied over a wider temperature range. For example, we find that the x = 0.003 sample exhibits a C/T which depends linearly on temperature for $T_{cB} < T < T_{cA}$ (see Fig. 3). Analysis of the data in Fig. 2 shows that the slope is independent of x, and the extrapolated T = 0 K intercept increases very weakly with x. From this we infer that the thermodynamic properties of the A-phase in $UPt_3$ are essentially unchanged by addition of Pd. Measurements of other temperature-dependent A-phase properties, such as transverse sound attenuation[7], may give greater insight into the nature of the unconventional superconductivity in this system.

Work was supported through Research Corporation grant RA0246 (MJG), NATO grant CRG960116 (MJG and AdV), and through the Dutch funding agency FOM.

# Figure Captions

Fig. 1.  Data for transition temperature $T_{cA}$ (solid symbols) and splitting $\Delta T_c$ (open symbols) versus Pd-content.  Circles are from this work, squares are from Ref. [1].  The lines are polynomial fits to the data.

Fig. 2.  Specific heat data.  Curves have the same normal-state value at $T_c$, but are offset for clarity.

Fig. 3.  Specific heat data for $x = 0.003$.  Solid circles are the data, and the line represents an idealized fit to the double superconducting transition using the criteria from Ref. [1].

FIGURE 1

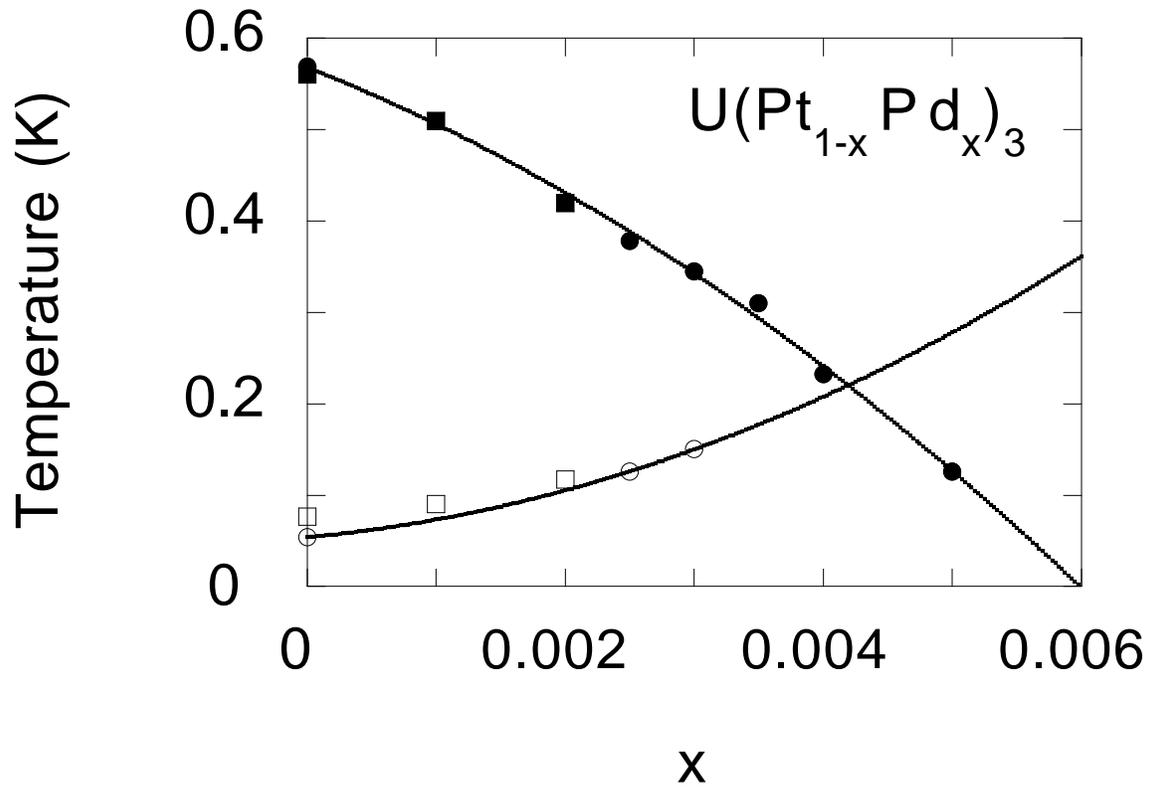

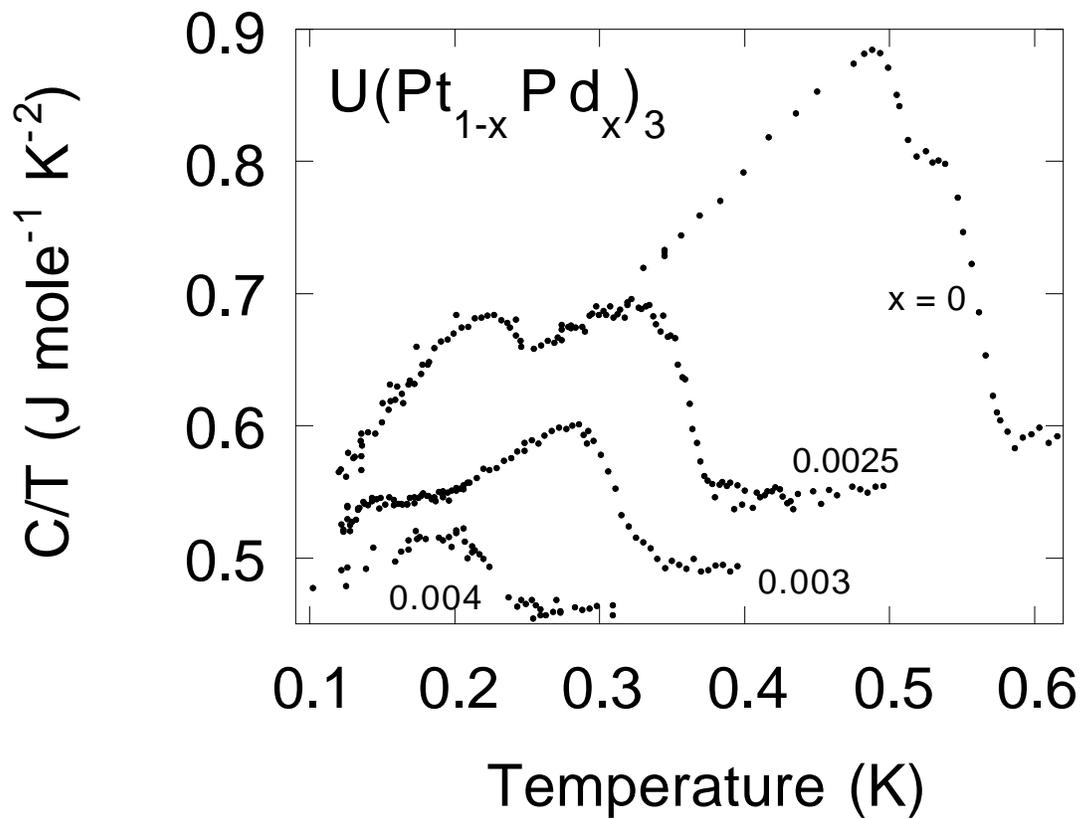

FIGURE 3

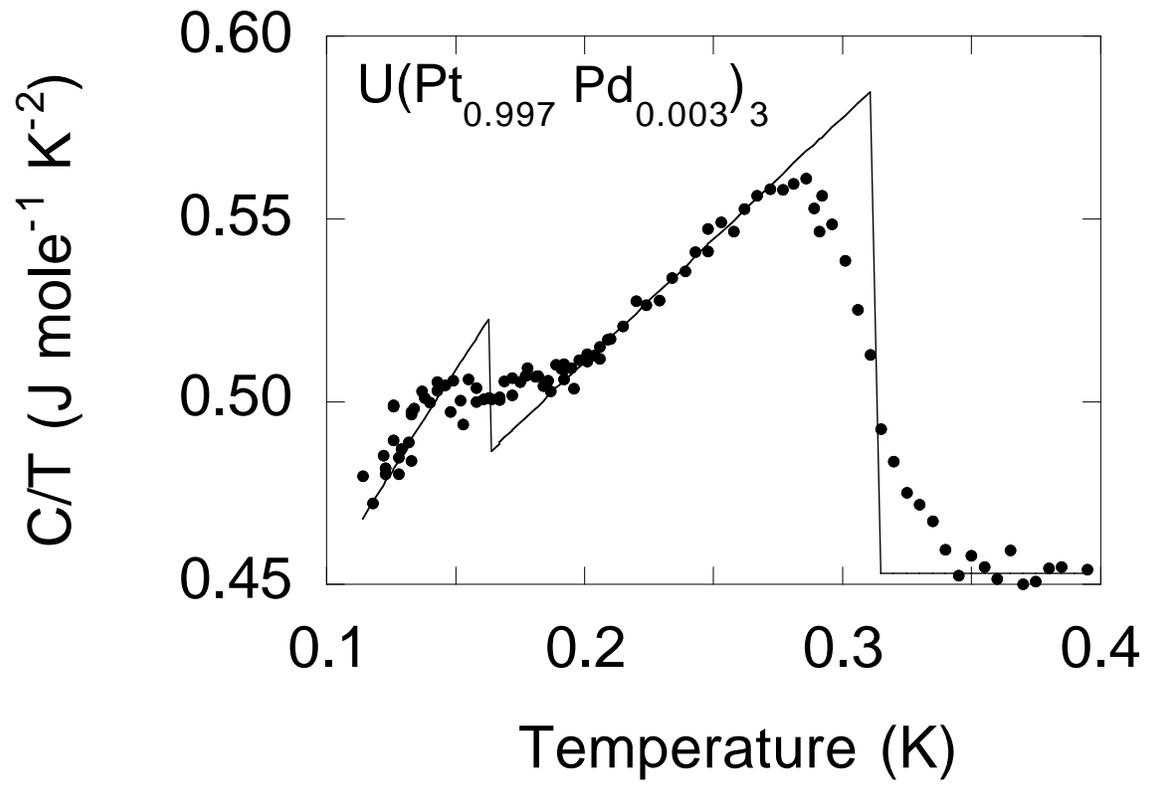